\documentclass[journal,twoside,web]{ieeecolor}
\usepackage{generic}
\usepackage{cite}
\usepackage{amsmath,amssymb,amsfonts}
\usepackage{algorithmic}
\usepackage{graphicx}
\usepackage{textcomp}
\usepackage{stfloats}
\usepackage{soul}

\usepackage{tabularx}

\usepackage{subcaption}
\usepackage[hidelinks]{hyperref}

\def\BibTeX{{\rm B\kern-.05em{\sc i\kern-.025em b}\kern-.08em
    T\kern-.1667em\lower.7ex\hbox{E}\kern-.125emX}}
    
\begin{document}
\title{KID-PPG: Knowledge Informed Deep Learning for Extracting Heart Rate from a Smartwatch}
\author{Christodoulos Kechris, Jonathan Dan, Jose Miranda, and David Atienza, \IEEEmembership{Fellow, IEEE}
\thanks{All authors are affiliated with the Embedded Systems Laboratory, EPFL, Switzerland}
\thanks{Corresponding author C.K. e-mail: christodoulos.kechris@epfl.ch}}

\maketitle

\begin{abstract}
Accurate extraction of heart rate from photoplethysmography (PPG) signals remains challenging due to motion artifacts and signal degradation. Although deep learning methods trained as a data-driven inference problem offer promising solutions, they often underutilize existing knowledge from the medical and signal processing community. In this paper, we address three shortcomings of deep learning models: motion artifact removal, degradation assessment, and physiologically plausible analysis of the PPG signal. We propose KID-PPG, a knowledge-informed deep learning model that integrates expert knowledge through adaptive linear filtering, deep probabilistic inference, and data augmentation. We evaluate KID-PPG on the PPGDalia dataset, achieving an average mean absolute error of 2.85 beats per minute, surpassing existing reproducible methods. Our results demonstrate a significant performance improvement in heart rate tracking through the incorporation of prior knowledge into deep learning models. This approach shows promise in enhancing various biomedical applications by incorporating existing expert knowledge in deep learning models. 
\end{abstract}

\begin{IEEEkeywords}
Photoplethysmograhy, Heart Rate, Motion Artifacts, Acceleration, Source Separation, Deep Learning, Knowledge Informed AI
\end{IEEEkeywords}

\section{Introduction}
\label{sec:introduction}

Photoplethysmography (PPG) is a non-invasive technique used to optically acquire the Blood Volume Pulse (BVP) \cite{allen2007photoplethysmography}. Its widespread adoption and ease of integration into wearable devices, especially smartwatches, have made PPG a popular choice for continuous and unobtrusive heart rate monitoring compared to electrocardiography (ECG). However, movement can introduce significant artifacts into PPG signals, complicating signal interpretation. These motion artifacts (MA) can overlap the actual BVP signal,  further complicating their removal \cite{biswas2019heart}. To address this challenge, numerous methods have been proposed to estimate the heart rate (HR) of PPG corrupted by MA. These methods generally fall into two categories: Signal Processing (SP) and Deep Learning (DL) approaches.

SP methods focus mainly on isolating the BVP component and minimizing the impact of MA \cite{biswas2019heart}. These methods often use a motion reference signal acquired from sensors such as accelerometers or gyroscopes to aid in MA removal \cite{lee2022adaptive, zhang2014heart, zhang2014troika, xu2019photoplethysmography, yang2018novel}. Acceleration and angular velocity are generally agreed to be effective references for periodic motion but offer limited correlation with PPG signals during random movements~\cite{lee2022adaptive}. Once the motion components are filtered out, HR is typically extracted from the filtered PPG signal by identifying its principal frequency component~\cite{schack2017computationally, salehizadeh2015novel, huang2020robust, zhou2020heart, lee2022adaptive, zhang2014heart, zhang2014troika, xu2019photoplethysmography, yang2018novel}, which is then attributed to the heart activity.

DL offers an alternative approach by combining filtering and HR estimation within a single model. Similarly to SP methods, DL methods have used acceleration as a reference signal for motion. In these models, both PPG and acceleration are given directly as inputs to the network and are fused by the model to produce a point estimate of HR \cite{reiss2019deep, kasnesis2022multi, burrello2021q} or an HR distribution \cite{bieri2023beliefppg, ray2022deeppulse}. The network is typically trained in a supervised manner using synchronized ECG-derived HR as ground truth labels. Several advanced DL techniques, including attention~\cite{kasnesis2022multi} and data augmentation~\cite{burrello2022improving}, have been used for PPG-based HR extraction.

Existing DL methods typically approach PPG-based HR estimation as a purely data-driven inference task, overlooking valuable prior knowledge from the medical and signal processing fields regarding BVP and PPG. Integrating task-specific prior knowledge into machine learning models has emerged as a promising strategy to improve explainability, robustness, and generalizability, particularly in scenarios with limited available data~\cite{von2021informed}. 

In this work, we explore the integration of prior knowledge into DL models for PPG-based HR inference. We identify failure cases of current DL models and propose three mechanisms to integrate prior knowledge into DL models effectively addressing these shortcomings. The resulting approach, which we term KID-PPG, represents a knowledge-informed DL-based HR inference model. To conduct our analysis, we take advantage of the publicly available PPGDalia dataset \cite{reiss2019deep}. Our code for the experiments is available here:  \url{https://github.com/esl-epfl/KID-PPG-Paper}. Through this investigation, we shed new light on the efficacy of DL models in processing PPG signals affected by MA and provide valuable insights into the recoverability of the BVP under challenging MA conditions. Our contributions include:
\begin{itemize}
    \item We identify three key factors contributing to erroneous HR estimations: DL models fail to separate MA from BVP, infer out-of-distribution HR samples, and estimate HR in sample affected by catastrophic MA.
    \item We address these limitations by incorporating prior knowledge into the DL workflow: explicitly defined MA separation task, guided probabilistic inference, and data augmentation.
    \item We design KID-PPG, a DL model for HR inference, achieving a mean absolute error (MAE) of 2.96 beats per minute (BPM) on PPGDalia. KID-PPG also provides estimates of uncertainty as a proxy for the assessment of the severity of the BVP artifact along with HR inference.
    \item We provide an open-source package to estimate HR from PPG and acceleration signals to the research community, which allows to repeat all our experiments and move forward the field of PPG analysis on embedded devices and wearables. The package is available here: \url{https://github.com/esl-epfl/KID-PPG}. 
\end{itemize}

\section{Methodology}
\label{sec:methods}
\begin{figure}[h]
    \centering
    \includegraphics[width=0.9\linewidth]{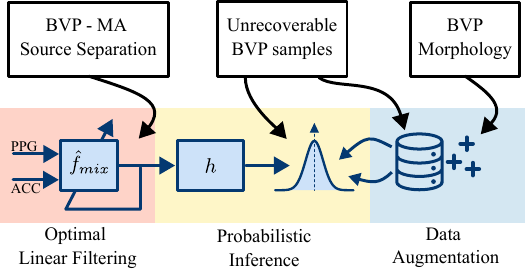}
    \caption{Knowledge-Informed deep-learning (DL) for heart-rate extraction (KID-PPG) incorporates prior knowledge on motion artifacts (MA), unrecoverable blood volume pulse (BVP) samples, and BVP morphology into DL models through three mechanisms: linear filtering, probabilistic inference, and data augmentation. The input of KID-PPG consists of a PPG signal along with an accelerometer signal. The linear filter ($\hat{f}_{mix}$) separates the BVP from MA to produce a filtered PPG signal, which serves as an input to the DL model ($h$). The model uses probabilistic inference to assess the degradation of the PPG signal and uses data augmentation to better characterize the BVP morphology.}
    \label{fig:knowledge_informed_idea_demo}
\end{figure}

We have identified the following three key points underutilized in existing DL models for HR extraction:
\begin{enumerate}
    \item Robust HR tracking requires MA removal.
    \item In some cases, BVP can be degraded by MA to the extent that it is unrecoverable.
    \item The BVP has specific morphology and characteristics.
\end{enumerate}This prior-knowledge is incorporated into our DL models through three mechanisms: Explicit Source Separation, Guided Probabilistic Inference and Data Augmentation. An overview of our methodology is presented in Fig. \ref{fig:knowledge_informed_idea_demo}.

\subsection{Motion Artifact Removal}
\label{sec:methods:motion_artefact_removal}
An inherent challenge in HR inference based on PPG is the mixing of MA components ($x_{MA}(t)$) with heart-related BVP ($x_{bvp}(t)$) through a mixing process denoted as $f_{mix}$:
\begin{equation} \label{eq:ppg_as_heart_and_motion_mix}
    x_{ppg}(t) = f_{mix}(x_{bvp}(t), x_{MA}(t), t)
\end{equation}
As is common practice, HR is inferred on 8-second windows with a 2-second overlap \cite{burrello2021q, kasnesis2022multi, reiss2019deep, ray2022deeppulse, bieri2023beliefppg, masinelli2021spare, zhang2014heart, zhang2014troika, song2021ppg, risso2021robust}. Hence, the i-th 8-second PPG sample is denoted as the $(N \times 1)$ vector $\textbf{x}_{ppg_i} = [x_{ppg}(t_i), x_{ppg}(t_i + \Delta t), ..., x_{ppg}(t_i + 8)]$, where $\Delta t = \frac{1}{f_s}$, $f_s$ is the sensor's sampling frequency, and $N = 8 sec \cdot f_s$ samples. Similarly, the corresponding BVP is denoted as $\textbf{x}_{bvp_i}$. The 3-axis acceleration (ACC) samples form the $(N \times 3)$ matrix $X_{acc_i} = [\textbf{x}_{acc_{x_i}}, \textbf{x}_{acc_{y_i}}, \textbf{x}_{acc_{z_i}}]$. Robust HR inference depends on the heartbeat-related $\textbf{x}_{bvp}$ \cite{zhang2014heart}. However, MA can significantly distort the morphology of BVP in the PPG observation (Eq. \ref{eq:ppg_as_heart_and_motion_mix}) \cite{zhang2014troika}. Many SP methods employ source separation to approximate the unmixing process $f_{mix}^{-1}$, allowing the HR inference module to rely on an approximation $\hat{x}_{bvp}(t) \approx x_{bvp}(t)$ \cite{zhang2014troika, zhang2014heart, masinelli2021spare, lee2022adaptive}, thus mitigating the effect of MA. 

In contrast, training a deep end-to-end HR estimator, $g(\cdot)$, with inputs $[\textbf{x}_{ppg_i}, X_{acc_i}]$, for the HR inference task, \cite{burrello2021q, kasnesis2022multi, reiss2019deep, ray2022deeppulse, bieri2023beliefppg, song2021ppg, risso2021robust}, does not guarantee modeling of the unmixing process, namely: 
\begin{equation}
\label{eq:unmixing_network}
    g = \hat{f}_{mix}^{-1} \circ h
\end{equation} where $h(\cdot)$ is an estimator of HR after component unmixing. Instead, the network may employ various fusion strategies $g = f_{fusion} \circ h^{\prime}$, where $h^{\prime}$ is an HR estimator after the two modalities have been fused by $f_{fusion}$. The convergence criterion for $f_{fusion}$ is the minimization of the HR loss, e.g. the MAE. Thus, this approach does not ensure BVP-based HR inference and may allow the network to learn spurious relations with the motion signals \cite{scimeca2021shortcut}.

Therefore, we propose the inclusion of an explicitly defined source separation task to disentangle the BVP component from MA. Without loss of generality, we model the motion artifact mixing $f_{mix}$ as a linear process \cite{foo2006computational, yang2018novel}, although other approaches are available \cite{ye2016combining, kim2017characterization}. Furthermore, we assume that $f_{mix}$ is a stationary process and that hand acceleration is a suitable reference signal for $x_{MA}(t)$ \cite{lee2022adaptive}. Hence, Eq. \ref{eq:ppg_as_heart_and_motion_mix} can be written as:
\begin{multline}\label{eq:linear_mixing_model}
    x_{ppg}(t) = x_{bvp}(t) \\
    + A_{mix} \ast [x_{acc_x}(t), x_{acc_y}(t), x_{acc_z}(t)] + noise(t)
\end{multline}
where $A_{mix}$ is a spatio-temporal mixing filter, $x_{acc_x}$, $x_{acc_y}$, $x_{acc_z}$ are the accelerations of the hand on the corresponding axis, and $\ast$ is the convolution operation. Separating the motion artifacts involves estimating the approximation $\hat{A}_{mix} \approx A_{mix}$. 

\begin{figure}[h]
    \centering
    \includegraphics[width=0.6\linewidth]{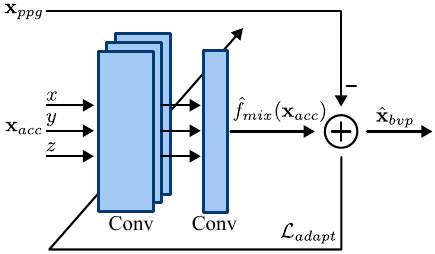}
    \caption{Linear model for separating the motion artifacts (MA) from the blood volume pulse (BVP) components in the PPG signal. A linear two-layer convolutional network takes the three axis of the accelerometer signal as an input ($\textbf{x}_{acc}$ along with the PPG ($\textbf{x}_{ppg}$) to produce a filtered PPG signal ($\hat{\textbf{x}}_{bvp}$).}
    \label{fig:adaptive_filtering}
\end{figure}

To estimate $\hat{A}_{mix}$ we use a linear two-layer convolutional network denoted as $\hat{f}_{mix}$ (Fig. \ref{fig:adaptive_filtering}). The first convolutional layer applies linear spatio-temporal filtering on the 3-channel accelerometer signals, while the second layer merges the three channels into one MA estimation. 

Similarly to \cite{yang2018novel}, we train $\hat{f}_{mix}$ in an unsupervised manner on pairs $ \left( X_{acc_i}, \textbf{x}_{ppg_i}\right)$ using an adaptive filter loss function, denoted as $\mathcal{L}_{adapt}$:

\begin{multline}\label{eq:unmixing_network_loss}
    \mathcal{L}_{adapt} = \\
    MSE \left( FFT\{\hat{f}_{mix}( X_{acc_i})\}, FFT\{ \textbf{x}_{ppg_i}\}\right)
\end{multline}
Here, $MSE$ represents the Mean Squared Error, defined as $\mathbb{E}[Error^2]$, and $FFT\{\cdot\}$ denotes the Fast Fourier Transform. Once the training converges, we use the resulting $\hat{f}_{mix}$ to estimate $\hat{\textbf{x}}_{bvp} \approx \textbf{x}_{bvp}$:

\begin{equation}\label{eq:unmixing_network_clean_bvp}
    \hat{\textbf{x}}_{bvp} = \textbf{x}_{ppg} - \hat{f}_{mix}(X_{acc})
\end{equation}

The estimated $\hat{\textbf{x}}_{bvp}$ is then inputted into a DL model, denoted as $h$ (Eq. \ref{eq:unmixing_network}).

\subsection{Temporal Attention Model}
\label{sec:methods:temporal_attention_model}

A convolutional neural network based on \cite{kasnesis2022multi} is used to extract an embedding $W_i$ for each sample $\hat{\textbf{x}}_{bvp_i}$.

HR temporal relationships are usually modeled as a smoothing filter during post-processing, reducing temporal granularity, or with LSTMs, increasing the model's complexity. We propose to model the temporal relationship between two consecutive samples as an attention operation. This approach allows us to consider the progression of the PPG signal embeddings while maintaining temporal granularity and computational simplicity. Let $E_{i - 1}, E_i$ denote the embeddings of two consecutive PPG frames. The multi-head temporal attention operation can be defined as \cite{vaswani2017attention, kasnesis2022multi}:
\begin{equation}
    Attention_{Temp}(E_i, E_{i - 1}) = softmax\left(\frac{E_{i}E_{i - 1}^T}{\sqrt{d}}\right)E_{i - 1}
\end{equation} where $d$ is the dimensionality of the embedding. Additionally, we incorporate a residual connection \cite{he2016deep, vaswani2017attention}: $E_i + Attention_{Temp}(E_i, E_{i - 1})$. 

\subsection{Guided Probabilistic Heart Rate Extraction}
\label{sec:methods:probabilistic_heart_rate_extraction}
\begin{figure}[h]
    \centering
    \includegraphics[width=0.4\textwidth]{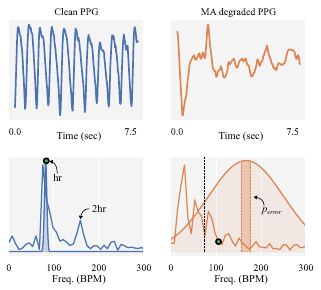}
    \caption{Probabilistic HR inference example on a clean (left) vs MA degraded PPG sample (right). The top row shows the raw PPG data. The bottom row shows the FFT of the PPG. The example is taken from S6 of PPGDalia. The green circles indicate the true ECG HR for the two samples. The probability density functions of $\mathcal{N}(\mu_{hr}, \sigma_{hr}^2)$ are overlayed on the frequency representations of the corresponding PPG sample. The error classification probability, $p_{error}$ is also illustrated. In the right sample the non-probabilistic point estimate HR inference of DL model is represented as a black vertical line (Q-PPG). }
    \label{fig:probabilistic_intuition_example}
\end{figure}

The DL models discussed in \cite{burrello2021q, reiss2019deep, kasnesis2022multi} and Sub-sections \ref{sec:methods:motion_artefact_removal} and \ref{sec:methods:temporal_attention_model} produce a point estimate of HR, implying that each sensor readout value contains a BVP component which can be isolated. However, in real-world conditions, the BVP component might be degraded beyond the point of reconstruction, leaving the MA component as the sole source of information in the PPG sample, Eq. \ref{eq:ppg_as_heart_and_motion_mix}. If $x_{bvp}(t)$ is not observed, the model performs HR extraction on irrelevant information, Eq. \ref{eq:ppg_as_heart_and_motion_mix}. This is illustrated in Fig. \ref{fig:probabilistic_intuition_example} where the BVP component is severely degraded, yet a point-estimate DL model continues to infer HR (Q-PPG \cite{burrello2021q}) .

To address this issue, we propose to design the model as a probability estimator of $HR$. Specifically, we choose the normal distribution parameterized as $HR \sim \mathcal{N}(\mu_{hr}, \sigma_{hr}^2)$. Inspired by \cite{ray2022deeppulse} and \cite{kendall2017uncertainties}, we model the heteroscedastic aleatoric uncertainty using a two-unit fully connected output to represent both $\mu_{hr}$ and $\sigma_{hr}$. If no heart-related information is available in the PPG sample, then any physiologically valid heart rate is possible, therefore, the HR estimator should produce a large uncertainty, as depicted in Fig. \ref{fig:probabilistic_intuition_example}.

The estimated HR distribution can also be used as an error classifier. We define the error classification probability as
\begin{multline}
    p_{error}(Thr \mid \textbf{x}_{ppg}, \mu_{hr}, \sigma_{hr}) \\
    = P( \mu_{hr} - Thr < HR < \mu_{hr} + Thr \mid \mu_{hr}, \sigma_{hr} ) \\
    = F_{\mu_{hr}, \sigma_{hr}}(\mu_{hr} + Thr) - F_{\mu_{hr}, \sigma_{hr}}(\mu_{hr} - Thr)
\end{multline} where $F_{\mu_{hr}, \sigma_{hr}}(x)$ is the cumulative distribution of the Gaussian distribution $\mathcal{N}(\mu_{hr}, \sigma_{hr}^2)$: $F_{\mu_{hr}, \sigma_{hr}}(x) = \Phi(\frac{x - \mu_{hr}}{\sigma_{hr}})$ and $Thr$ is the threshold above which the error is considered significant. This error classifier predicts a high probability of the HR estimation being untrustworthy (error $\geq Thr$) when $p_{error} \geq CL$, where $CL$ is the required confidence level. As is usual in classifiers we set the confidence level to 0.5, although it can be freely selected to stricter values depending on the application.

\begin{figure}[h]
    \centering
    \includegraphics[width=0.4\textwidth]{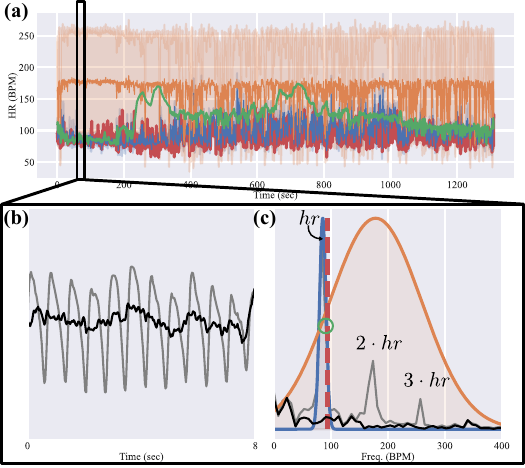}
    \caption{Inference after filtering the BVP component out of the PPG. Example taken from S6 of PPGDalia. \textbf{(a)} Inferences across the entire session from Q-PPG (red), probabilistic (blue) and guided probabilistic (orange). For the probabilistic models, the range of one standard deviation is also presented. \textbf{(b)} Sample example with initially clean PPG (grey) and synthetically degraded (black). \textbf{(c)} Frequency domain representation of the example sample and HR inferences from the three models. True HR is presented with a green circle. Both Q-PPG and probabilistic models estimate HR close to the ground truth, indicating potential learned shortcuts since there is no heart rate information in the signals. In contrast, the guided probabilistic model estimates a large standard deviation identifying the lack of relevant information.}
    \label{fig:probabilistic_augmentation_demo}
\end{figure}

To improve the robustness of our error classifier, we guide the network to base its HR inference on the BVP component, as relying on HR-inference loss only can lead to spurious behavior. Fig. {\ref{fig:probabilistic_augmentation_demo}} illustrates this with a synthetic example generated from PPGDalia (S6). Here we have manually removed the BVP component, by applying a bandstop filter with cut-off frequencies around $HR$, $2 \cdot HR$ and $3 \cdot HR$, yet both the probabilistic and the Q-PPG models continue to infer HR  (MAE 4.11 BPM and 3.33 BPM accordingly). The probabilistic model is overconfident, with the $p_{error}(Thr = 10BPM)$ classifier dropping only $7.44\%$ of the samples. 

We tackle this limitation by guiding the training process to map PPG samples with severe BVP degradation to a normal distribution with a high standard deviation. To do this we generate realistic samples in which the BVP component is extremely degraded. It has been empirically observed that in PPG recordings, BVP information is mainly located at the HR frequency, and it second and third harmonics \cite{cho2012preliminary, masinelli2021spare}. Therefore, for each sample $(\textbf{x}_{ppg_i}, hr_i)$ in the original training dataset, we filter $\textbf{x}_{ppg_i}$ by band-stopping (Finite Input Response with 81 taps) the frequencies around $hr_i,\ 2hr_i,\ 3hr_i$ (cutoff from $i \cdot hr - 2.5 BPM$ to $ i \cdot hr + 2.5 BPM$), forming a new signal $\textbf{x}_{noise_i}$ (Fig. \ref{fig:probabilistic_augmentation_demo} b, c). Using the original $\textbf{x}_{ppg_i}$ as the seed for $\textbf{x}_{noise_i}$ helps maintain realism. Since we need a ground truth label for the supervised training task, we create a random HR label: $hr_{noise_i} \sim Uniform(40, 300)$. The lowest HR value is set to account for slower HR \cite{rijnbeek2014normal} and the highest reflects the theoretical maximum human HR \cite{chhabra2012mouse}. The pair $(\textbf{x}_{noise_i}, hr_{noise_i})$ is then added to the set of auxiliary adversarial examples. $50\%$ of the adversarial examples set is randomly sampled to form the \textit{adversarial-examples-subset}.

\begin{figure}[h]
    \centering
    \includegraphics[width=0.4\textwidth]{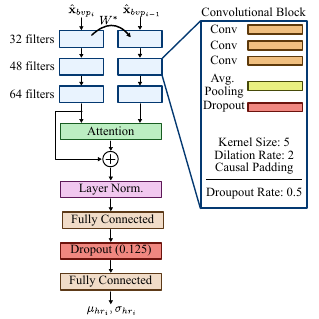}
    \caption{KID-PPG network architecture. $W^*$ indicates weight sharing between the convolution blocks for the $\hat{\textbf{x}}_{bvp_i}$ and $\hat{\textbf{x}}_{bvp_{i-1}}$ branches.}
    \label{fig:kid_ppg_network_architecture}
\end{figure}

In summary, by adopting a probabilistic approach to HR estimation and guiding the network to focus on the BVP component, we aim to improve the reliability of HR inference from PPG signals. The full KID-PPG network configuration is presented Fig. \ref{fig:kid_ppg_network_architecture}.

\subsection{Data Augmentation}
\label{sec:methods:data_augmentation}
We propose a data augmentation scheme to address the limited number of high HR samples in the available datasets, similarly to~\cite{burrello2022improving}. After removing the MA with {$f_{mix}$}, from the samples of the training set, we generate synthetic PPG waveforms corresponding to higher heart rate frequencies, forming the \textit{high-heart-rate} training subset. The following procedure is followed: 
\begin{enumerate}
    \item Locate the 8-sec. samples designated as clean PPG for which the main frequency component is close to the ground truth HR.
    \item Artificially speed up the sample by x2.
    \item Discard samples with $HR \ge 300 BPM$ (\cite{chhabra2012mouse}).
\end{enumerate}

The original training dataset, the \textit{high-heart-rate} and \textit{adversarial-examples-subset} subsets are then merged into the final training set.

\section{Experimental Setup}

For our experiments we use the publicly available PPGDalia data \cite{reiss2019deep}. This dataset comprises synchronized ECG and wrist-worn PPG and acceleration recordings from 15 subjects, with approximately two-hour recording sessions per subject. The PPG and acceleration signals were collected using the Empatica E4 wristband. During the two-hour session, the subjects underwent diverse activities to simulate daily life conditions: resting, ascending/descending stairs, playing table soccer, cycling, driving a car, having lunch, walking and working in an office. Between the activities there is a transition period. A systematic temporal shift between PPG and ACC was identified and manually corrected. The ECG-based HR provided in the dataset serves as the ground truth. We adopt the leave-one-subject-out cross-validation procedure proposed in \cite{reiss2019deep} for all experiments. 

Additionally, to showcase KID-PPG's generalizability we validate a pre-trained model on the WESAD dataset \cite{schmidt2018introducing}. WESAD is comprised of recordings from 15 subjects. The session for each subject involves activities evoking various stress levels. More details on the activities can be found in the original publication \cite{schmidt2018introducing}. We select a model trained during the PPGDalia cross-validation process and validate it on all the subjects of WESAD without any further training.

For the MA removal step, we train an adaptive model separately for each subject and activity given that converging $\hat{f}_{mix}$ using $\mathcal{L}_{adapt}$, Eq. \ref{eq:unmixing_network_loss}, requires stationarity of the $f_{mix}$ process. We employ The MAE loss function and the stochastic gradient descent (SGD) to converge the model ($lr = 1e-7, momentum = 1e-2$).

To evaluate the MA removal step, we employ the Q-PPG \cite{burrello2021q}, Attention Model \cite{kasnesis2022multi} and Temporal Attention model (Sub-section \ref{sec:methods:temporal_attention_model}) as HR inference models, $h$ Eq. \ref{eq:unmixing_network}. For the Q-PPG and Attention models we used the hyperparameters reported in \cite{burrello2021q} and \cite{kasnesis2022multi}, respectively. The Temporal Attention model was trained using Adam optimizer ($lr = 0.0005, \beta_1 = 0.9, \beta_2 = 0.999, \epsilon = 1e-08$).

The probabilistic models where trained using the negative log-likelihood (NLL) loss and the same training strategy used for the non-probabilistic ones. NLL is also used to evaluate their performance. Furthermore, we use the True Positive Rate (TPR) and the F1 score to evaluate the classifier $p_{error}$ for its ability to correctly identify untrustworthy samples. We consider a positive classification when the error classifier predicts a high probability of error, $p_{error}(Thr) \leq 0.5$, the error threshold was arbitrarily selected at $Thr = 10BPM$.

A summary of all experiments is presented in Table \ref{tab:experimental_setup_summary}.

\begin{table}[h]
    \centering
    \begin{tabularx}{\linewidth}{>{\hsize=0.65\hsize}X|
                                  >{\hsize=1.35\hsize}X
                                }
        \textbf{PPGDalia} & \\
        \hline
        \hline
        \textbf{Base models} & \\
        • Q-PPG & Introduced in A. Burrello et al.~\cite{burrello2021q}. \\
        • Attention & Introduced in P. Kasnesis et al.~\cite{kasnesis2022multi}.  \\
         \hline
        \textbf{Adaptive} & \\
        • Adaptive + Q-PPG & Sub-section~\ref{sec:methods:motion_artefact_removal} \\
        • Adaptive + Attention  & \\
        \hline
        \textbf{Data augmentation} &\\
        • Adaptive + Attention \newline+ HR Aug. & Sub-section \ref{sec:methods:data_augmentation}\\
        \hline
        \textbf{Probabilistic} &\\
        • Probabilistic model & Adaptive + Attention model adapted for probabilistic inference (Sub-section \ref{sec:methods:probabilistic_heart_rate_extraction}).\\[5mm]
        • Probabilistic Temporal Attention model & Probabilistic model with temporal attention (Sub-section \ref{sec:methods:temporal_attention_model}).\\[5mm]
        • KID-PPG  & Probabilistic model with temporal attention, guided probabilistic training and High-HR augmentation (Sub-section 
        \ref{sec:methods:probabilistic_heart_rate_extraction}).\\
        \textbf{WESAD} & \\
        \hline
        \hline
        • KID-PPG  & A pretrained model from the PPGDalia experiments is validated on WESAD without any further training.\\
    \end{tabularx}
    \caption{Summary of evaluated models.}
    \label{tab:experimental_setup_summary}
\end{table}

\section{Results}
\label{sec:results}

\subsection{Results on Motion Artifact Removal}
\begin{figure*}[t]
    \centering
    \includegraphics[width = 0.9\textwidth]{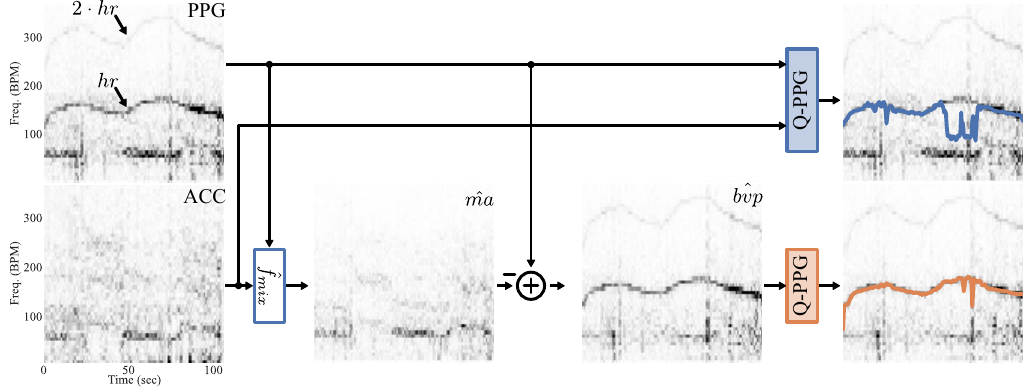}
    \caption{Estimating HR on stairs activity of S6 with Q-PPG (top row - blue) and with Adaptive filtering and Q-PPG (bottom - orange). The original Q-PPG (blue) takes the PPG and accelerometer (ACC) as input.  Adaptive + Q-PPG (orange) takes the output ($\hat{bvp})$ of the adaptive filter ($\hat{f}_{mix}$) as input. } 
    \label{fig:S6_results_demo_qppg_vs_adaptive}
\end{figure*}

The adaptive linear filter effectively removes MA that are linearly coupled with acceleration signals. An illustrative example is provided in Fig. \ref{fig:S6_results_demo_qppg_vs_adaptive} for subjects S6 during an episode where the subject is walking up and down stairs. The DL model alone fails to decouple motion information, resulting in significant errors (activity MAE 13.66 BPM). In contrast, the Adaptive + DL Model combination substantially reduces the error (activity MAE 2.74 BPM). In particular, during periods when Q-PPG loses track of the HR, the BVP signal, located at $hr$ and $2 \cdot hr$, remains discernible and thus the loss of HR tracking is not attributed to severe degradation of BVP due to MA.

\begin{figure}[h]
    \centering
    \includegraphics{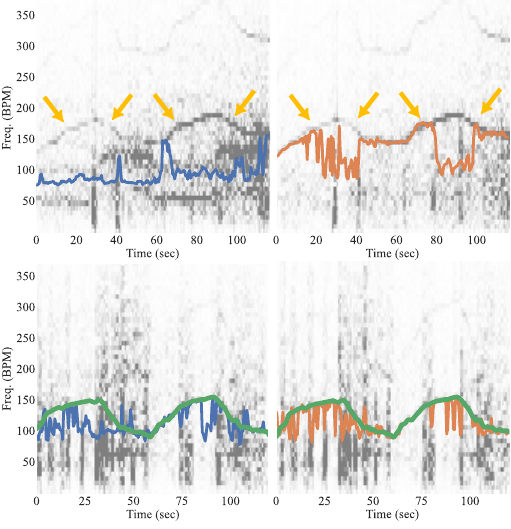}
    \caption{Effect of the MA Removal for the stairs activity. \textbf{Top:} Subjects S5 with Q-PPG (blue) and Adaptive + Q-PPG (orange). The ground truth heart rate is clearly visible in the spectrogram as a black line as indicated by the yellow arrow. \textbf{Bottom:} Subject S9. Ground truth HR is represented in green.}
    \label{fig:ma_removal_demo_S5_S9}
\end{figure}

Similarly, for S5 (Fig. \ref{fig:ma_removal_demo_S5_S9}), the BVP component is visually identifiable, yet the DL model fails to disentangle it from MA (activity MAE 69.96 BPM). Our filtering approach effectively isolates the BVP, resulting in lower HR inference error (activity MAE 24.96 BPM). However, during certain periods, 20-40sec. and 80-100sec., the DL model loses track of the HR. We discuss this further in Sub-section \ref{sec:results:high_hr_augmentation}.

For S9 (Fig. \ref{fig:ma_removal_demo_S5_S9}) the situation differs slightly. MA removal still outperforms the DL model, with Q-PPG stairs activity MAE 16.91 BPM vs. MAE 10.37 BPM for the Adaptive + Q-PPG. However, in this case, MA degradation is more severe, and the MA component is not linearly coupled with ACC. Consequently, the DL model sometimes loses track of the HR even after the MA removal step. This limitation is addressed by the error classifier.

The overall per-subject results are presented in Table \ref{tab:performance_of_point_estimate_hr_models}. Both Q-PPG and the Attention models perform better when data are preprocessed by the adaptive step, with MAE 4.81 BPM (Q-PPG) vs. 3.98 (Q-PPG with adaptive step) and MAE 4.44 vs. 3.70 for the Attention model. Notably, adding a preprocessing step to Q-PPG outperforms the original Attention model. For all but one subject, S9, the best performing model is obtained by employing the adaptive preprocessing step (Table \ref{tab:performance_of_point_estimate_hr_models}). 

\begin{table*}[h!]
\centering
\begin{tabularx}{\linewidth}{>{\hsize=5.6\hsize}X|
                             >{\hsize=0.71\hsize}X
                             >{\hsize=0.71\hsize}X
                             >{\hsize=0.71\hsize}X
                             >{\hsize=0.71\hsize}X
                             >{\hsize=0.71\hsize}X
                             >{\hsize=0.71\hsize}X
                             >{\hsize=0.71\hsize}X
                             >{\hsize=0.71\hsize}X
                             >{\hsize=0.71\hsize}X
                             >{\hsize=0.71\hsize}X
                             >{\hsize=0.71\hsize}X
                             >{\hsize=0.71\hsize}X
                             >{\hsize=0.71\hsize}X
                             >{\hsize=0.71\hsize}X
                             >{\hsize=0.71\hsize}X|
                             >{\hsize=0.75\hsize}X
                            }
& S1   & S2   & S3   & S4   & S5    & S6    & S7   & S8    & S9    & S10  & S11   & S12   & S13  & S14  & S15  & Avg \\
\hline
\hline
\multicolumn{17}{l}{\textbf{Signal Processing}}\\
\hline
SpaMaPlus \cite{reiss2019deep} & 8.86 & 9.67 & 6.4  & 14.1 & 24.06 & 11.34 & 6.31 & 11.25 & 16.04 & 6.17 & 15.15 & 12.03 & 8.50  & 7.76 & 8.29 & 11.06   \\
TAPIR \cite{huang2020robust} & 4.50  & 4.50  & 3.20  & 6.00    & 5.00     & 3.40   & 2.80  & 6.30   & 8.00     & 2.90  & 5.10   & 4.70  & 3.10  & 5.00    & 4.10 & 4.57 \\
CurToSS \cite{zhou2020heart}  & 5.40  & 4.30  & 3.00   & 8.00    & \textbf{2.20}   & 2.80   & 3.30  & 8.50   & 12.60  & 3.60  & 3.60   & 6.10   & 3.00 & 5.50  & 3.70  & 5.04       \\
\hline
\multicolumn{17}{l}{\textbf{Deep Learning}}\\
\hline
DeepPPG \cite{reiss2019deep} & 7.73 & 6.74 & 4.03 & 5.90  & 18.51 & 12.88 & 3.91 & 10.87 & 8.79  & 4.03 & 9.22  & 9.35  & 4.29 & 4.37 & 4.17 & 7.65 \\
NAS-PPG \cite{song2021ppg} & 5.46 & 5.01 & 3.74 & 6.48 & 12.68 & 10.52 & 3.31 & 8.07  & 7.91  & 3.29 & 7.05  & 6.76  & 3.84 & 4.85 & 3.57 & 6.16 \\
Q-PPG \cite{burrello2021q} & 4.29 & 3.62 & 2.44 & 5.73 & 10.33 & 5.26  & 2.00 & 7.09  & 8.6   & 3.09 & 4.99  & 6.25  & 1.92 & 3.02 & 3.55 & 4.81 \\
TEMPONet \cite{burrello2022improving} & 4.37 & 3.74 & 2.43 & 5.49 & 13.48 & 5.71  & 2.23 & 7.86  & 8.94  & 3.32 & 5.34  & 7.71  & 2.03 & 2.94 & 3.58 & 5.27 \\
Augmentation \cite{burrello2022improving} & 4.97 & 4.34 & 2.39 & 6.14 & 9.41  & 3.63  & 2.23 & 9.14  & 10.98 & 3.4  & 5.27  & 7.64  & 2.05 & 2.84 & 3.61 & 5.20 \\
TimePPG \cite{risso2021robust} & 4.51 & 3.37 & 2.33 & 5.25 & 14.68 & 4.76  & 2.37 & 8.04  & 8.75  & 3.30  & 5.19  & 8.08  & 2.29 & 3.02 & 3.49 & 5.30 \\
Q-PPG \cite{burrello2021q} & 4.29 & 3.62 & 2.44 & 5.73 & 10.33 & 5.26  & 2.00    & 7.09  & 8.60   & 3.09 & 4.99  & 6.25  & 1.92 & 3.02 & 3.55 & 4.81 \\
Attention \cite{kasnesis2022multi} & 4.75 & 3.31 & 2.22 & 5.25 & 7.43  & 4.22  & 2.28 & 8.93  & 6.95  & 2.93 & 3.98  & 6.57  & 1.70  & 3.22 & 2.88 & 4.44 \\
\hline
\multicolumn{17}{l}{\textbf{Ours}}\\
\hline
Adaptive + Q-PPG & 3.80 (6.77) & 3.50 (6.99) & 2.07 (4.07) & 5.18 (7.67) & 5.76 (14.92) & 3.38 (7.24) & 1.59 (2.15) & 6.99 (9.79)  & 8.94 (10.86) & 2.75 (5.80) & 3.37 (7.40)  & 5.61 (8.85) & 1.49 (2.72) & 2.45 (4.68) & 2.80 (5.87) & 3.98 (7.05) \\
Adaptive + Attention & 4.55 (9.22) & 3.12 (5.97) & 1.88 (3.54) & 5.27 (7.73) & 4.00 (12.14) & \textbf{2.54} (5.83) & \textbf{1.28} (1.85) & 7.20 (10.41) & 8.60 (11.07) & 2.61 (6.03) & 3.06 (6.39) & 4.91 (9.10) & 1.50 (2.59) & \textbf{2.28} (4.99) & 2.64 (6.31) & 3.70 (6.88)\\
Adaptive + Attention + HR Aug & 4.27 (7.68) & 3.46 (8.83) & 2.07 (3.78) & 5.61 (9.78) & 3.01 (9.41) & 2.74 (6.99) & 1.39 (2.31) & 7.13 (9.65) & 9.53 (14.50) & 2.77 (6.36) & 3.58 (7.76) & 4.52 (7.16) & \textbf{1.48} (2.75) & 2.48 (5.32) & 2.84 (5.82) & 3.79 (7.21)\\
\hline
\multicolumn{17}{l}{\textbf{Ours - Probabilistic}}\\
\hline
KID-PPG ($Thr = 10BPM$) & \textbf{2.99} (4.21) & \textbf{2.54} (4.40) & \textbf{1.84} (2.59) & \textbf{3.83} (5.48) & 2.8 (6.52) & 2.56 (5.33) & 1.50 (1.84) & \textbf{5.16} (7.03) & \textbf{4.67} (6.92) & \textbf{2.33} (3.77) & \textbf{2.90} (4.80) & \textbf{3.73} (5.13) & 1.72 (2.28) & 2.33 (3.61) & \textbf{1.91} (2.79) & \textbf{2.85} (4.45)\\
KID-PPG (\% retention) & 79.89 & 87.08 & 92.73 & 66.76 & 88.35 & 93.12 & 99.33 & 72.51 & 48.32 & 86.63 & 86.12 & 79.25 & 99.16 & 92.57 & 83.55 & 83.69
\end{tabularx}

\caption{Mean Absolute Error and Standard Deviation of the Absolute Error (in parenthesis)} performance of point-estimate HR and probabilistic (after retention) models on the PPGDalia dataset. All values are in Beats-per-Minute with the exception of the retention percentages which are presented for KID-PPG (probabilistic model).
\label{tab:performance_of_point_estimate_hr_models}

\end{table*}

\subsection{High-HR Augmentation} 
\label{sec:results:high_hr_augmentation}

\begin{figure}[h]
    \centering

    \includegraphics[width=0.45\textwidth]{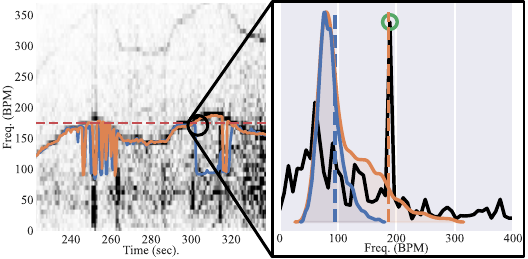}

    \caption{Effect of out-of-distribution samples on the model inference. \textbf{(a)} HR inference of the model with (orange) and without (blue) High-HR augmentation. The maximum HR value in the training set is presented with a red dashed line. \textbf{(b)} Example sample in the frequency domain along with model prediction and ground truth HR (green circle). The HR distributions of the non-augmented and High-HR augmented training sets are also analyzed in the corresponding colors. }
    \label{fig:S5_high_hr_results}
\end{figure}

Training with High-HR augmentation results in a higher MAE (3.79 BPM) compared to not augmenting (3.70 BPM). Subjects S1, S5, S8 and 13 benefited from the augmentation (Table \ref{tab:performance_of_point_estimate_hr_models}). High-HR augmentation teaches the model to infer HR in a broader range than that of the original dataset (Fig. \ref{fig:S5_high_hr_results}). In Fig. \ref{fig:S5_high_hr_results} the HR inference error for S5 during Stairs activity decreases MAE from 25.33BPM to 6.73BPM. Consequently, for the augmented model, there is a wider range of MA components that can be mistakenly inferred as HR. Additionally, the expected HR is now artificially inflated, leading to higher errors in considerably degraded samples in which HR inference reflects the HR distribution encoded in the model weights . As a result, the model's overall MAE is increased. This limitation is addressed by the probabilistic mechanism (Sub-section \ref{sec:results:guided_probabilistic_inference}).

\subsection{Guided Probabilistic Inference}
\label{sec:results:guided_probabilistic_inference}

The NLL losses for the probabilistic models are summarized in Table \ref{tab:nll_results_summary}. Our proposed methodology, KID-PPG, significantly outperforms BeliefPPG \cite{bieri2023beliefppg} (NLL 4.78) and reduces NLL by approximately 40\%. KID-PPG also manages to generalize to a never-before-seen dataset (WESAD) achieving an NLL of 3.15, outperforming BeliefPPG (NLL 4.7 - 32\% reduction).

\begin{table}[h]
    \centering
    \begin{tabular}{l|c}
        \textbf{Method}                         & \textbf{Mean Negative Log-Likelihood} \\
        \hline
        \hline
        BeliefPPG \cite{bieri2023beliefppg}     &         4.78                          \\
       \hline
        Adapt + Attention + Prob                &         3.25                          \\
        Adapt + Temp. Attention + Prob          &         3.08                          \\
        KID-PPG                                 & \textbf{2.99}                         \\
    \end{tabular}
    \caption{Summarized NLL results for the proposed probabilistic models.}
    \label{tab:nll_results_summary}
\end{table}

\begin{figure}
    \centering
    \includegraphics{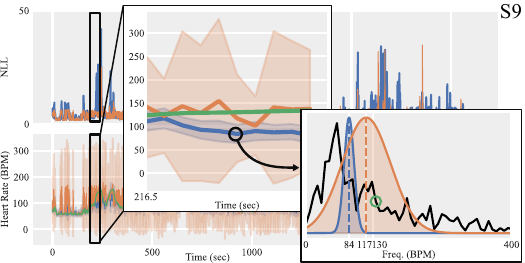}
    \caption{Example of probabilistic output of augmented (orange) vs non-augmented model (blue) for subject S9 and ground truth HR (green circle). For the example, its frequency domain representation verifies that the PPG signal is considerably corrupted. The augmented probabilistic model manages to identify the lack of BVP content, presenting a high standard deviation, in contrast to the non-augmented one.}
    \label{fig:probabilistic_example_S9}
\end{figure}

Although the non-augmented and augmented probabilistic temporal attention models achieve similar mean NLL, they differ in their ability to infer untrustworthy samples. The guided probabilistic model dropped $98.58\%$ of the samples in the example in Fig. \ref{fig:probabilistic_augmentation_demo}. A real data example is presented in Fig. \ref{fig:probabilistic_example_S9}.  The non-augmented NLL presents "spikes" or outliers corresponding to overconfident but wrong HR estimations, with the highlighted sample showing a 46.30 BPM Absolute Error with 9.72 BPM STD. Adversarial augmentation acts as a probabilistic regularization, reducing the networks' overconfidence (12.98 BPM Absolute Error with 46.92 BPM STD). 

\begin{figure}[h]
    \centering
    \includegraphics[width = 0.49\textwidth]{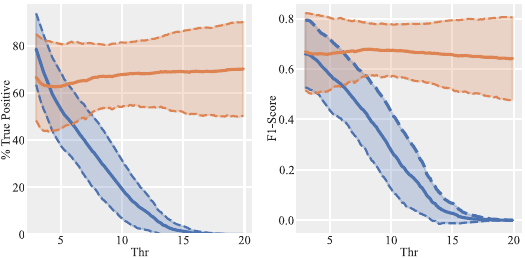}
    \caption{TPR and F1-score for KID-PPG (Guided Probabilistic training and High-HR augmentation), orange, vs Unguided probabilistic model (Unguided probabilistic training and no High-HR augmentation), blue.}
    \label{fig:overall_probabilistic_classification_performance}
\end{figure}

The overall performance of the error classifiers for KID-PPG and the unguided probabilistic models, TPR and F1-Score, is depicted in Fig. \ref{fig:overall_probabilistic_classification_performance}. Selecting a high threshold ($Thr = 10BPM$), KID-PPG achieves an average MAE of 2.85 BPM.

\section{Discussion}

Our analysis provides novel insights into the challenges of PPG-based HR inference, particularly within the PPGDalia dataset. The difficulties encountered by DL model to accurately estimate HR have often been attributed to out-of-distribution high HR samples \cite{reiss2019deep, risso2021robust}. Specifically, S5 has been highlighted as a particularly challenging subject \cite{risso2021robust, burrello2022improving, reiss2019deep}. Our findings emphasize three key factors contributing to erroneous HR estimations: DL models fail to separate MA from BVP, out-of-distribution HR samples and catastrophic MA.

\textbf{DL models encounter difficulties in separating linear MA}. The introduction of an explicit linear MA separation filter notably improved model performance, enabling HR inference in cases where the DL model alone had failed. This finding suggests a possible physics-based explanation, as hinted at by Lee et al. \cite{lee2022adaptive}, who empirically linked wrist-worn PPG motion artifact components to the physical forces acting on the hand during physical activities.

\textbf{Out-of-distribution HR samples} have posed challenges, particularly exemplified by subject S5,  frequently cited in the literature for its higher HR samples. Previous studies have reported improvement through data augmentation, particularly in S5 and S6~\cite{burrello2022improving} . In our experiments, mainly subjects S1 and S5 benefited from increased HR. A comparison between S5 and S6 reveals differing underlying causes of high errors, with S6 primarily affected by the PPG - MA coupling, and S5 impacted by both MA and out-of-distribution samples. 

\textbf{Catastrophic MA} remains a significant challenge, particularly evident in subjects S8 and S9, which consistently exhibit high MAE. Despite various proposed solutions, no one has achieved a low MAE for these two subjects (Table {\ref{tab:performance_of_point_estimate_hr_models}}). Furthermore, the rejection of a significant portion of their samples by KID-PPG (27.48\% and 53.16\% of their samples respectively $\left(Thr = 10BPM\right)$) indicates severe BVP corruption. This underscores the need for a probabilistic HR estimator and a robust error classifier based on the morphology of the BVP components.

While our work has demonstrated the efficacy of explicitly defining the source-separation task, real-world deployment scenarios, which further explore the assumption of stationarity of $f_{mix}$, Eq. \ref{eq:unmixing_network_clean_bvp} are required. Deploying this solution requires identifying when $\hat{f}_{mix}$ needs to be updated, potentially through an activity/context recognition module. In addition, other implementations of the source-separation task warrant investigation.

In integrating BVP-morphology-related prior knowledge into KID-PPG, we have proposed a method for creating realistic PPG signals with extreme BVP degradation. Exploring alternative approaches to generate synthetic MA-affected PPG samples and expanding the understanding of MA dynamics with the BVP could inform the design of more realistic augmentation techniques. Evaluating MA degradation presents an additional challenge, as HR-inference accuracy may not fully reflect the level of degradation, for example Fig.{\ref{fig:probabilistic_augmentation_demo}}c. Addressing these challenges will be crucial to advancing the robustness and reliability of PPG-based HR inference systems.

\section{Conclusion}
In this study, we have introduced a novel method, called KID-PPG, to enhance DL models with expert knowledge integration for HR inference from PPG inputs (combined with a motion reference signal). In particular, we have proposed three main knowledge integration mechanisms: Adaptive Linear Filtering, Guided Probabilistic Inference, and Data Augmentation. Adaptive Filtering removes linear MA, enabling the DL model to infer HR accurately even in segments with high MA, thus strongly reducing MAE compared to existing DL models. Probabilistic inference allows the model to assess the BVP degradation in input signals due to artifacts. Using the proposed error classifier, the model can selectively retain samples with a high probability of maintaining BVP information, improving overall inference accuracy. Our guided probabilistic training strategy, tailored to the morphology of BVP, substantially improved the robustness of the error classifier. Data Augmentation extended the range of HR that can be reliably inferred by the model, further enhancing its performance. This work highlights the important benefits of integrating expert knowledge into DL models. All in all, KID-PPG achieves an overall MAE of 2.85 BPM ($Thr = 10BPM$) on PPGDalia, outperforming all reproducible methods. Our findings demonstrated the efficacy of our approach in advancing HR inference from PPG signals, paving the way for improved healthcare monitoring and diagnostics.

\section*{Acknowledgment}

This research was partially supported by the PEDESITE Swiss NSF Sinergia
project (GA No. CRSII5\textunderscore 193813 / 1), the RESoRT project (GA No. REG-19-019) from the Botnar Foundation, and by the Wyss Center for Bio and Neuro Engineering through funding for ESL-EPFL in the Non-invasive Neuromodulation of Subcortical Structures project of the Ligthhouse Partnership Agreement with EPFL.


\end{document}